\documentstyle[twoside,fleqn,espcrc2]{article}

\newcommand{\AmS}{{\protect\the\textfont2
  A\kern-.1667em\lower.5ex\hbox{M}\kern-.125emS}}

\def\beq{\begin{equation}}
\def\eeq{\end{equation}}
\def\bea{\begin{eqnarray}}
\def\eea{\end{eqnarray}}
\def\bq{\begin{quote}}
\def\eq{\end{quote}}
\def\ve{\vert}
\def\nnb{\nonumber}
\def\ga{\left(}
\def\dr{\right)}
\def\aga{\left\{}
\def\adr{\right\}}

\def\rar{\rightarrow}
\def\nnb{\nonumber}
\def\la{\langle}
\def\ra{\rangle}
\def\nin{\noindent}
\title{HEAVY QUARKS FROM QCD SPECTRAL SUM RULES}
\author{ Stephan Narison\address{
Theoretical Physics Division, CERN
CH - 1211 Geneva 23, Switzerland\\
and\\
Laboratoire de Physique Math\'ematique,
Universit\'e de Montpellier II\\
Place Eug\`ene Bataillon,
34095 - Montpellier Cedex 05, France}}
\begin{document}
\pagestyle{plain}
\begin{abstract}
\noindent
We summarize the recent developments on the extraction of the
dynamical properties of the heavy quarks from QCD spectral sum rules.
\end{abstract}
\maketitle
\section{Introduction} \par
We have been living with QCD spectral sum rules (QSSR)
(or QCD sum rules, or ITEP sum rules, or hadronic sum rules...)
for 15 years, within the impressive ability
of the method for describing the complex phenomena of hadronic
physics with the few universal ``fundamental" parameters of the QCD
Lagrangian
(QCD coupling $\alpha_s$, quark masses
and  vacuum condensates built from the quarks
and/or gluon fields, which parametrize the
non-perturbative phenomena). The approach might be very close to the
lattice calculations as it also uses the first principles of QCD, but
unlike the case of the lattice, which is based  on
sophisticated numerical simulations, QSSR is
quite simple as it is a semi-analytic approach based
on a semiperturbative expansion and Feynman graph
techniques implemented in an Operator Product Expansion (OPE),
where  the condensates contribute as higher-dimension
operators. The QCD information is transmitted to the data
via a dispersion relation obeyed by the hadronic correlators,
in such a way that
in this approach, one can {\it really control and in some sense localize}
the origin of the numbers obtained from the analysis. With this
simplicity, QSSR can describe in an elegant way the complexity
of the hadron phenomena, without waiting for a complete understanding
of the confinement problem.

\nin
One can fairly say that QCD spectral sum rules already started, before
QCD, at the time of current algebra, in 1960, when different {\it ad hoc}
superconvergence sum rules, especially the Weinberg and
Das--Mathur--Okubo sum rules, were proposed but they came
under control only with the advent of QCD \cite{FLO}. However,
the main flow comes from the classic paper of
Shifman--Vainshtein--Zakharov
\cite{SVZ} (hereafter referred to as SVZ), which goes beyond
the {\it na\"\i ve} perturbation
theory thanks to the inclusion of the vacuum condensate effects in
the OPE (more
details and more complete discussions of QSSR and its various
applications to hadron physics can be found, for instance, in
\cite{SNB}).

\nin
In this talk, I shall present aspects of QSSR in the analysis of the
properties of heavy flavours. As I am limited in space-time
(an extended and updated version of this talk will be
published elsewhere \cite{SNH}),
I cannot cover in detail here all QSSR applications to the
heavy-quark physics. I will only focus on the following
topics, which I think are important in the development of the
understanding of the heavy-quark properties in connection with the
progress done recently in the heavy quark effective theory (HQET) and in
Lattice calculations:

\nin
-- heavy-quark masses,

\nin
-- pseudoscalar decay constants and the bag parameter $B_B$,

\nin
-- heavy-to-light semileptonic and radiative decay form factors,

\nin
--$SU(3)$ breaking in $\bar{B}/D \rar Kl\bar{\nu}$ and determination
of $V_{cd}/V_{cs}$,

\nin
-- slope of the Isgur-Wise (IW) function and determination of $V_{cb}$,

\nin
-- properties of hybrids and $B_c$-like mesons.

 \section{ QCD spectral sum rules}
In order to illustrate the QSSR method in a pedagogical way, let us consider
the two-point correlator:
\bea
\Pi^{\mu\nu}_b &\equiv& i \int d^4x ~e^{iqx} \
\la 0\vert {\cal T}
J^{\mu}_b(x)
\ga J^{\nu}_b(o)\dr ^\dagger \vert 0 \ra \nnb \\
&=& -\ga g^{\mu\nu}q^2-q^\mu q^\nu \dr \Pi_b(q^2,M^2_b),
\eea
where $J^{\mu}_b(x) \equiv \bar b \gamma^\mu b (x)$ is the local vector
current of the $b$-quark. The correlator obeys the well-known
K\"allen--Lehmann dispersion relation:
\beq
\Pi_b(q^2,M^2_b) = \int_{4M^2_b}^{\infty} \frac{dt}{t-q^2-i\epsilon}
{}~\frac{1}{\pi}~\mbox{Im}  \Pi_b(t) ~ + ...,
\eeq
which expresses in a clear way the {\it duality} between the spectral
function Im$ \ \Pi_b(t)$, which can be measured experimentally, as here
it is related to the $e^+e^-$ into $\Upsilon$-like states total
cross-section, while $\Pi_b(q^2,M^2_b)$ can be calculated directly in
QCD, even at $q^2=0$,
thanks to the fact that $M^2_b-q^2 \gg \Lambda^2$.
The QSSR is an improvement on the previous
dispersion relation.

\nin
On the QCD side, such an improvement is achieved by adding
to the usual perturbative expression of the correlator,
the non-perturbative contributions as parametrized by the vacuum
condensates of higher and higher dimensions in the OPE \cite{SVZ}:
\bea
\Pi_b (q^2,M^2_b)
&\simeq& \sum_{D=0,2,4,...}\frac{1}{\ga M^2_b-q^2 \dr^{D/2}} \nnb
\\
&.&\sum_{dim O=D} C^{(J)}(q^2,M^2_b,\mu)\la O(\mu)\ra, \nnb \\
&&
\eea
where $\mu$ is an arbitrary scale that separates the long- and
short-distance dynamics; $C^{(J)}$ are the Wilson coefficients calculable
in perturbative QCD by means of Feynman diagrams techniques:
$D=0$
corresponds to the case of the na\"\i ve perturbative contribution;
$\la O \ra$ are
the non-perturbative condensates built
 from the quarks or/and gluon
fields. For $D=4$, the condensates that can be formed are the
quark $M_i \la\bar \psi \psi \ra$ and gluon $\la\alpha_s G^2 \ra$
ones; for
$D=5$, one can have the mixed quark-gluon condensate $\la\bar \psi \sigma_
{\mu\nu}\lambda^a/2 G^{\mu\nu}_a \psi \ra$, while for $D=6$ one has, for
instance, the
triple gluon $gf_{abc}\la G^aG^bG^c \ra$ and the four-quark
$\alpha_s \la \bar \psi \Gamma_1 \psi \bar \psi \Gamma_2 \psi \ra$, where
$\Gamma_i$ are generic notations for any Dirac and colour matrices. The
validity of this expansion has been understood formally, using renormalon
techniques (IR renormalon ambiguity is absorbed into the definitions
of the condensates)
\cite{MUELLER} and by building  renormalization-invariant
combinations of the condensates (Appendix of \cite{PICH} and references
therein). The SVZ expansion is phenomenologically confirmed from the
unexpected
accurate determination of the QCD coupling $\alpha_s$ from semi-inclusive
tau decays \cite{PICH,ALFA}. In the present case of heavy-heavy
correlators the OPE is much simpler, as one can show \cite{HEAVY,BC}
that the heavy-quark condensate
effects can be included into those of the gluon condensates, so
that, up to $D\leq 6$, only the $G^2$ and $G^3$ condensates appear in
the OPE. Indeed, SVZ have, originally, exploited this feature for their
first estimate of the gluon condensate value.

\nin
For the phenomenological side, the improvement comes from the uses of
either a finite number of derivatives and finite values of $q^2$
(moment sum rules):
\bea
{\cal M}^{(n)}& \equiv& \frac{1}{n!}\frac{\partial^n \Pi_b(q^2)}
{\ga \partial q^2\dr^n} \Bigg{\vert} _{q^2=0} \nnb \\
&=& \int_{4M^2_b}^{\infty} \frac{dt}{t^{n+1}}
{}~\frac{1}{\pi}~ \mbox{Im}  \Pi_b (t),
\eea
or an infinite number of derivatives and infinite values of $q^2$, but
keeping their ratio fixed as $\tau \equiv n/q^2$
(Laplace or Borel or exponential sum rules):
\beq
{\cal L}(\tau,M^2_b)
= \int_{4M^2_b}^{\infty} {dt}~\mbox{exp}(-t\tau)
{}~\frac{1}{\pi}~\mbox{Im} \Pi_b(t).
\eeq
There also exist non-relativistic versions of these two sum rules,
which are convenient quantities to work with in the large-quark-mass
limit. In these cases, one introduces non-relativistic
variables $E$ and $\tau_N$:
\beq
t \equiv (E+M_b)^2 \ \ \ \ \mbox{and} \ \ \ \  \tau_N = 4M_b\tau .
\eeq
In the previous sum rules,
the gain comes from the weight factors, which enhance the
contribution of the lowest ground-state meson to the spectral integral.
Therefore, the simple duality ansatz parametrization:
\bea
&&``\mbox{one resonance}"\delta(t-M^2_R)
 \ + \nnb \\
&& ``\mbox{QCD continuum}" \Theta (t-t_c),
\eea
of the spectral function,
gives a very good description of the spectral integral, where the
resonance enters via its coupling to the quark current. In the case
of the $\Upsilon$, this coupling can be defined as:
\beq
\la 0\vert \bar b\gamma^\mu b \vert \Upsilon \ra =
 \sqrt{2} \frac{M^2_\Upsilon}
{2\gamma_\Upsilon}.
\eeq
The previous
feature
has been tested in the light-quark channel from the $e^+e^- \rar$
$I=1$ hadron data and in the heavy-quark ones from the
$e^+e^- \rar \Upsilon$ or $\psi$ data, within a good
accuracy.
To the previous sum rules, one can also add the ratios:
\beq
{\cal R}^{(n)} \equiv \frac{{\cal M}^{(n)}}{{\cal M}^{(n+1)}}~~~~~
\mbox{and}~~~~~
{\cal R}_\tau \equiv -\frac{d}{d\tau} \log {{\cal L}},
\eeq
and their finite energy sum rule (FESR) variants, in order to fix
the squared mass of the ground state.
\nin
In principle, the pairs $(n,t_c)$, $(\tau,t_c)$ are free external
parameters in the analysis, so that the optimal result should be
insensitive to their variations. Stability criteria, which are equivalent
to the variational method, state that the best results should
be obtained at the minimas or at the inflexion points in $n$ or $\tau$,
while stability in $t_c$ is useful to control the sensitivity of the
result in the changes of $t_c$ values. To these stability criteria are
added constraints from local duality FESRs, which
correlate the $t_c$ value to those of the ground state mass and
coupling \cite{FESR}. Stability criteria have also been tested in
models such as the
harmonic oscillator, where the exact and approximate
solutions are known \cite{BELL}. The {\it most conservative
optimization criteria}, which include various types of optimizations
in the literature, are the following: the
$optimal$ result is obtained in the region, starting at
the beginning of $\tau / n$ stability (this corresponds in most
of the cases to the so-called plateau often discussed in the literature,
but in my opinion, the interpretation of this nice plateau as a
sign of a good continuum model is not sufficient, in the sense
that the flatness of the
 curve extends in the uninteresting high-energy region where the
properties of the ground state are lost),
until the beginning of the $t_c$
stability, where the value of $t_c$ more or less
corresponds to the one fixed by FESR duality constraints.
The earlier {\it sum rule window} introduced by SVZ, stating that the
optimal result should be in the region where both the non-perturbative
and continuum contributions are {\it small}, is included in the previous
region.
 Indeed, at the stability
point, we have an equilibrium between the continuum and non-perturbative
contributions, which are both small,
while the OPE is still convergent  at this point.

\section{The heavy-quark-mass values}
Here, we will summarize the recent results obtained in \cite{SNM}, where
an improvement and an update of the existing results have been done,
with the emphasis that the apparent discrepancy encountered in the
literature is mainly due to the different values of $\alpha_s$ used by
various authors. Using the $world~average$
value $\alpha_s(M_Z)=0.118 \pm 0.006$ \cite{BETHKE}, the $first$
determination of the running mass to two loops, from the $\Psi$ and
$\Upsilon$ systems, is:
\bea
\overline{m}_c(M^{PT2}_c) &=& (1.23 ^{+ 0.02}_{-0.04}\pm 0.03
)~ \mbox{GeV} \nnb \\
\overline{m}_b(M^{PT2}_b)
&=& (4.23 ^{+ 0.03}_{-0.04}\pm 0.02)~ \mbox{GeV},
\eea
where the errors are respectively due to $\alpha_s$ and to the gluon
condensate. One can transform this result into the $perturbative$
pole mass and obtain, to two-loop accuracy:
\bea
M^{PT2}_c & = & (1.42 \pm 0.03
)~ \mbox{GeV} \nnb \\
M^{PT2}_b
&= &  (4.62 \pm 0.02)~ \mbox{GeV}.
\eea
It is informative to compare these values with the ones from the pole
masses from non-relativistic sum rules to two loops:
\bea
M^{NR}_c & =& (1.45^{+ 0.04}_{-0.03} \pm 0.03
)~ \mbox{GeV} \nnb \\
M^{NR}_b
& =&  (4.69 _{-0.01}^{+0.02}\pm 0.02)~ \mbox{GeV},
\eea
where one may interpret the small mass difference,
 less than 70 MeV as the
size of the renormalon effect into the pole mass. A similar comparison
can be done at three-loop accuracy. One obtains:
\bea
M^{PT3}_c  & =&  (1.62 \pm 0.07 \pm 0.03
)~ \mbox{GeV} \nnb \\
M^{PT3}_b
&=&  (4.87 \pm 0.05 \pm 0.02)~ \mbox{GeV},
\eea
to be compared with
 the $dressed~mass$:
\beq
  M_b^{nr}= (4.94\pm 0.10 \pm 0.03)~\mbox{ GeV},
\eeq
obtained from
a non-relativistic Balmer formula based on a $\bar bb $ Coulomb
potential and including higher order $\alpha_s
$-corrections \cite{YND}. Here,
one still has the 70 MeV mass difference, which reinforces
our interpretation that it is due to the renormalon effect (for
an exposition of this effect, see e.g. \cite{BENEKE}). One can also
use the previous results, in order to deduce the mass-difference
between the $b$ and $c$ (non)-relativistic pole masses:
\beq
M_b-M_c = (3.22 \pm 0.03)~ \mbox{GeV},
\eeq
in good agreement with potential model expectations \cite{RICH}.

\nin
An extension of the previous analysis to the case of the $B$ and $B^*$
mesons leads to the value $M_b^{PT2}=(4.63 \pm 0.08)$ GeV, in good
agreement with the previous results. The meson-quark mass difference
has been also directly estimated in the large mass limit.
By combining
the result from HQET \cite{NEU} with the one from the
full QCD spectral
sum rules \cite{SNA,SN4}, one can deduce the $weighted ~average$:
\bea
\bar\Lambda \equiv \delta M_b^{\infty} &\equiv&
(M_B-M^{NR}_b)_{\infty}\nnb \\
& = &(0.58 \pm 0.05)~\mbox{GeV},
\eea
of the quark-meson mass difference,
which is in agreement with the previous findings, but less accurate.

\nin
Using the previous result in (10) and the expression of the running
mass to two-loops, one also obtains at 1 GeV:
\bea
\overline{m}_c(1~\mbox{GeV}) &=& (1.46 ^{+ 0.09}_{-0.05}\pm 0.03
)~ \mbox{GeV} \nnb \\
\overline{m}_b(1~\mbox{GeV})
&=& (6.37 ^{+ 0.64}_{-0.39}\pm 0.07)~ \mbox{GeV},
\eea
By combining the previous value of the running $b$-quark mass
 with the $s$-quark one evaluated at 1 GeV, which we take from
 $\overline{m}_s$(1 GeV)= 150 MeV
 \cite{SN1} until 230 MeV \cite{GASSER} ,
 one obtains the scale-independent ratio:
\beq
\overline {m}_b/\overline {m}_s \simeq 33.5 \pm 7.6,
\eeq
a result of great interest for model-building and GUT-phenomenology.
\section{The pseudoscalar decay constants and the bag parameter $B_B$}
The decay constants $f_P$ of a pseudoscalar meson $P$ are defined as:
\beq
(m_q+M_Q)\la 0\vert \bar q (i\gamma_5)Q \vert P\ra
 \equiv \sqrt{2} M^2_P f_P,
\eeq
where in this normalization $f_\pi = 93.3$ MeV.
A {\it rigorous}
upper bound on these couplings can be derived from the
second-lowest superconvergent moment:
\beq
{\cal M}^{(2)} \equiv \frac{1}{2!}\frac{\partial^2 \Psi_5(q^2)}
{\ga \partial q^2\dr^2} \Bigg{\vert} _{q^2=0},
\eeq
where $\Psi_5$ is the two-point correlator associated to the pseudoscalar
current. Using the positivity of the higher-state contributions to the
spectral function, one can deduce \cite{SNZ}:
\beq
f_P \leq \frac{M_P}{4\pi} \aga 1+ 3 \frac{m_q}{M_Q}+
0.751 \bar{\alpha}_s+... \adr,
\eeq
where one should not misinterpret the mass-dependence in this
expression compared to the one expected from heavy-quark symmetry.
Applying this result to the $D$ meson, one obtains:
\beq
f_D \leq 2.14 f_\pi .
\eeq
Although
presumably quite weak, this bound, when combined with the recent
determination to two loops \cite{SN2}:
\beq
\frac{f_{D_s}}{f_D} \simeq (1.15 \pm 0.04)f_\pi ,
\eeq
implies
\beq
f_{D_s} \leq (2.46 \pm 0.09)f_\pi ,
\eeq
which is useful for a comparison with the recent measurement of $f_{D_s}$
from WA75: $f_{D_s} \simeq (1.76 \pm 0.52)f_\pi$ and from CLEO:
$f_{D_s} \simeq (2.61 \pm 0.49)f_\pi$.
One cannot push, however, the uses
of the moments to higher $n$ values in this $D$ channel, in order to
minimize the continuum contribution to the sum rule with the aim to
derive an estimate of the decay constant because the QCD series
will not converge at higher $n$ values.
In the $D$ channel, the most appropriate sum rule is the
Laplace sum rule. The results from different groups are consistent
for a given value of the $c$-quark mass. Using the table in
\cite{SN2} and the value of the perturbative pole mass obtained
previously, one obtains to two loops:
\bea
f_D &\simeq& (1.35 \pm 0.04\pm 0.06)f_\pi~~~~~~~ \Rightarrow  \nnb \\
f_{D_s} &\simeq& (1.55 \pm 0.10)f_\pi .
\eea
For the $B$ meson, one can either work with the Laplace, the
moments or
their non-relativistic variants. Given the previous value of $M_b$, these
different methods give consistent values of $f_B$, though the one
from the non-relativistic sum rule is very inaccurate due to the huge
effect of the radiative corrections in this method. The best value comes
from the Laplace sum rule; from the table in \cite{SN2}, one
obtains:
\beq
f_B \simeq (1.49\pm 0.06\pm 0.05)f_\pi ,
\eeq
while \cite{SN2}:
\beq
 \frac{f_{B_s }}{f_B}\simeq 1.16 \pm 0.04,
\eeq
where the most accurate estimate comes from the ``relativistic" Laplace
sum rule. One could notice, since the {\it first} result
$f_B \simeq f_D$ of \cite{SN3}, a large violation of the
scaling law expected from heavy-quark symmetry. Indeed,
 this is due to the large 1/$M_b$-correction
 found from the HQET sum rule \cite{NEU} and from the one in full
QCD \cite{SN4,SNA}:
\bea
f_B \sqrt{M_b} &\simeq& (0.42 \pm 0.07)~\mbox{GeV}^{3/2}\nnb \\
&.&\aga 1-\frac{
(0.88\pm
0.18)~\mbox{GeV}}{M_b}\adr,
\eea
which is due to the meson-mass gap $\delta M \equiv M_B-M_b$ \cite{NEU}
and to the continuum energy $E_c$ \cite{SN4,ZAL} ($E_c \simeq
\frac{3}{4} \delta M $ \cite{SNA}):
\beq
f_B \sqrt{M_b} \simeq \frac{1}{\pi}E_c^{3/2}\aga 1-\frac{\delta M}{M_b}
-\frac{3}{2}\frac{E_c}{M_b}+...\adr.
\eeq
The apparent disagreement among different existing
QSSR numerical results in the literature is due mainly to the
different values of the quark masses used because the decay constants
are very sensitive to that quantity as shown explicitly in \cite{SN2}.

\nin
Finally, let me mention that we have also tested the validity
of the vacuum saturation $B_B=1$ of the bag parameter, using a
sum rule analysis of the four-quark two-point correlator to two loops
\cite{PIVO}. We
found that the radiative corrections are quite small. Under some
 physically reasonable assumptions for the spectral function, we found
that the vacuum saturation estimate is only violated by about $15\%$,
giving:
\beq
 B_B \simeq 1 \pm 0.15.
\eeq
By combining this result with the one for $f_B$, we deduce:
\beq
f_B\sqrt{B_B} \simeq  (197\pm 18)~{\mbox MeV},
\eeq
if we use the normalization $f_\pi= 132$ MeV, which is $\sqrt 2$
times the one defined in (18),
in excellent agreement with the
present lattice calculations \cite{PENE}.
\section{Heavy-to-light semileptonic and radiative decay form factors}
One can extend the analysis done for the two-point correlator to the
more complicated case of three-point function, in order to study the form
factors related to the $B\rar K^*\gamma$ and $B\rar \rho/\pi$ semileptonic
decays. In so doing, one can consider the generic three-point function:
\bea
V(p,p',q^2)&\equiv &-\int d^4x~ d^4y ~e^{i(p'x-py)} \nnb \\
&&~\la 0|{\cal T}
 J_L(x)O(0)J^{\dagger}_B(y)|0\ra ,
\eea
where $J_L, ~J_B$ are the currents of the light and $B$ mesons; $O$
 is the
weak operator specific for each process (penguin for the $K^* \gamma$,
weak current for the semileptonic); $q \equiv p-p'$.
The vertex obeys the double dispersion
relation :
\bea
V(p^2,p'^2) &\simeq&
\int_{M_b^2}^{\infty} \frac{ds}{s-p^2-i\epsilon} \nnb \\
&&\int_{m_L^2}^{\infty} \frac{ds'}{s'-p'^2-i\epsilon}
{}~\frac{1}{\pi^2}~ \mbox{Im}V(s,s')
\nnb \\
\eea
As usual, the QCD part enters in the LHS of the sum rule, while the
experimental observables can be introduced through the spectral function
after the introduction of the intermediate states. The improvement of the
dispersion relation can be done in the way discussed
previously for the two-point
function. In the case of the heavy-to-light transition,
 the only possible
improvement with a good $M_b$ behaviour
at large $M_b$ (convergence of the QCD series) is the so-called
hybrid sum rule (HSR) corresponding to the uses of
the moments for the heavy-quark
channel and to the Laplace for the light one \cite{SNA,SN5}:
\bea
{\cal H} (n, \tau') &=&\frac{1}{\pi^2}
\int_{M^2_b}^\infty \frac{ds}{s^{n+1}} \nnb \\
&&\int_0^\infty ds'~e^{-\tau' s'}~\mbox{Im}V(s,s').
\eea
We have
studied analytically the different form factors entering the previous
processes \cite{SN6}. They are defined as:
\bea
 \la\rho(p')\ve \bar u \gamma_\mu (1-\gamma_5) b \ve B(p)\ra
=(M_B+M_\rho)A_1
\epsilon^*_\mu - \nnb \\
 \frac{A_2}{M_B+M_\rho}\epsilon^*p'(p+p')_\mu
+\frac{2V}{M_B+M_\rho} \epsilon_{\mu \nu \rho \sigma}p^\rho p'^\sigma ,
  \nnb \\
 \la\pi(p')\ve
\bar u\gamma_\mu b\ve B(p)\ra = f_+(p+p')_{\mu}+f_-(p-p')_\mu ,
 \nnb \\
\eea
and:
\bea
&&\la \rho(p')\ve \bar s \sigma_{\mu \nu}\ga
\frac{1+\gamma_5}{2}\dr q^\nu b\ve B(p)\ra \nnb \\
&&=i\epsilon_{\mu \nu \rho \sigma}\epsilon^{*\nu}p^\rho p'^\sigma
F^{B\rar\rho}_1 +
\nnb \\
&& \aga \epsilon^*_\mu(M^2_B-M^2_{\rho})-\epsilon^*q(p+p')_{\mu}
\adr \frac{F^{B\rar \rho}_1}{2}. \nnb \\
&&
\eea
We found that they are dominated, for $M_b \rar \infty$, by the
effect of the light-quark condensate, which dictates the $M_b$ behaviour
of the form factors to be
typically of the form:
\beq
F(0) \sim \frac{\la \bar dd \ra}{f_B}\aga 1+\frac{{\cal I}_F}
{M^2_b}\adr,
\eeq
where ${\cal I}_F$ is the integral from the perturbative triangle
graph, which is constant as $t'^2_cE_c/\la \bar dd \ra$ ($t'_c$ and
$E_c$ are the continuum thresholds of the light and $b$ quarks)
for large values of $M_b$. It
indicates that at $q^2=0$ and to leading order in $1/M_b$,
all form factors behave like $\sqrt{M_b}$,
although, in most cases, the coefficient of the $1/M^2_b$ term is large.
The
study of the $q^2$ behaviours of the form factors shows
 that, with the
exception of the $A_1$ form factor, their $q^2$ dependence
is only due to the non-leading (in $1/M_b$) perturbative graph, so that
for $M_b \rar \infty$,
these  form factors remain constant from $q^2=0$ to $q^2_{max}$.
The resulting $M_b$ behaviour at $q^2_{max}$ is the one expected from the
heavy quark symmetry. The numerical
effect of this $q^2$-dependence at finite values of $M_b$ is a polynomial in
$q^2$ (which can be resummed),
which mimics  quite well the usual pole parametrization for a pole mass
of about 6--7 GeV. The situation for the $A_1$ is drastically
different from
the other ones, as here the Wilson coefficient of the $\la \bar dd \ra$
condensate contains a $q^2$ dependence with a $wrong$ sign and reads:
\beq
A_1(q^2) \sim \frac{\la \bar dd \ra}{f_B}\aga 1-\frac{q^2}{M^2_b}
\adr ,
\eeq
which, for $q^2_{max} \equiv (M_B-M_\rho)^2$, gives the expected behaviour:
\beq
A_1(q^2_{max}) \sim \frac{1}{\sqrt{M_b}}.
\eeq
It can be noticed that the $q^2$ dependence of $A_1$
is in complete contradiction
with the pole behaviour due to its wrong sign. This result explains
the numerical analysis of \cite{BALL2}. It
is urgent and important to test this feature experimentally.
It should be finally noticed that owing
 to the overall $1/f_B$ factor, all
form factors have a large $1/M_b$ correction.

\nin
In the numerical analysis, we obtain at $q^2=0$, the value of the $B\rar
K^*\gamma$ form factor:
\bea
F_1^{B\rar \rho } &\simeq& 0.27 \pm 0.03,\nnb \\
\frac{F_1^{B\rar K^*}}{F_1^{B\rar \rho}}&\simeq& 1.14 \pm 0.02,
\eea
which leads to the branching ratio $(4.5\pm 1.1)\times 10^{-5}$, in perfect
agreement with the CLEO data and with the estimate in \cite{ALI}.
One should also notice that, in this case,
the coefficient of the $1/M^2_b$ correction is very large,
which makes dangerous the extrapolation of the
$c$-quark results to higher values of the
quark mass. This extrapolation is often done in some
lattice calculations.

\nin
For the semileptonic decays, QSSR
give a good determination of the ratios of
the form factors with the values \cite{SN5}:
\bea
&&\frac{A_2(0)}{A_1(0)} \simeq \frac{V(0)}{A_1(0)}
\simeq 1.11 \pm 0.01    \nnb  \\
&&\frac{A_1(0)}{F_1^{B\rar \rho}(0)} \simeq 1.18 \pm 0.06    \nnb \\
&&\frac{A_1(0)}{f_+(0)} \simeq 1.40 \pm 0.06.
\eea
Combining these results with the ``world average" value of $f_+(0)=
0.25 \pm 0.02$ and the one of $F_1^{B\rar \rho}(0)$, one can deduce the
rate and polarization:
\bea
&&\Gamma_\pi \simeq (4.3\pm 0.7)
|V_{ub}|^2 \times 10^{12}~\mbox{s}^{-1} \nnb \\
&&\Gamma_\rho /\Gamma_\pi \simeq 0.9 \pm 0.2    \nnb \\
&&\Gamma_+ /\Gamma_- \simeq 0.20\pm 0.01   \nnb \\
&&\alpha \equiv 2\frac{\Gamma_L}{\Gamma_T}-1 \simeq -(0.60 \pm 0.01).
\eea
These results are much more precise than
the ones from a direct estimate of the absolute values of the form
factors due to the cancellation of systematic errors in the ratios. They
indicate that, we are on the way to reach $V_{ub}$ with a good accuracy
from the exclusive modes.
Also
here, mainly because of the non-pole behaviour of $A^B_1$,
the ratio between the widths into $\rho$ and into
$\pi$ is about 1, while in different pole models,
 it ranges from 3 to 10.
For the
asymmetry, one obtains a large negative value of $\alpha$, contrary to
the case of the pole models.
\section{$SU(3)$ breaking in $\bar B/D \rar Kl\bar\nu$ and
   determination of $V_{cd}/V_{cs}$}
We extend the previous analysis for the estimate of the $SU(3)$ breaking
in the ratio of the form factors:
\beq
 R_P\equiv f_{+}^{P\rar K}(0) /f_{+}^{P\rar \pi}(0),
\eeq
where $P\equiv \bar B,\pi$. Its analytic expression is given in
 \cite{SN8}, which leads to the numerical result:
\beq
R_B = 1.007 \pm 0.020 ~~~~~~~~~~R_D=1.102\pm 0.007,
\eeq
where one should notice that for $M_b \rar \infty$, the SU(3) breaking
vanishes, while its size at finite mass is typically of the same order
as the one of $f_{D_s}$ or of the $B\rar K^*\gamma$ discussed before.
What is more surprising is the fact that using the previous value of $R_D
$ with the present value of CLEO data \cite{CLEO}:
\beq
\frac{Br(D^+\rar \pi^0 l\nu)}
{Br(D^+\rar \bar{K}^0 l\nu)}= (8.5 \pm 2.7 \pm 1.4)\%,
\eeq
one deduces:
\beq
V_{cd}/V_{cs}=0.322\pm 0.056,
\eeq
which is much larger than the value $0.226\pm 0.005$ derived from the
unitarity of the CKM matrix. This can mean either that the CLEO data are
wrong (recall that MARKIII data \cite{MARK}
would imply a value $0.25 \pm 0.15$, in
agreement with unitarity, but less accurate), or that the unitarity
constraint is in trouble. It is difficult to see how the QSSR result
is wrong as other predictions derived in the
same way (see e.g $f_{B_s}$ and $ F^{B\rar K^*}_1)$ agree successfully
with results from alternative approaches.
\section{Slope of the Isgur--Wise function and determination of
$V_{cb}$}
Let me now
discuss  the slope of the Isgur--Wise function. Taron--de Rafael
\cite{TARON} have exploited
the analyticity of the elastic $b$-number form factor
$F$ defined as:
\beq
\la B(p')|\bar b \gamma^\mu b |B(b)\ra =(p+p')^\mu F(q^2),
\eeq
which is normalized as $F(0)=1$
in the large mass limit $M_B \simeq M_D$. Using the positivity of the
vector spectral function and a mapping in order to get a bound on the
slope of $F$
outside the physical cut, they obtained a rigorous but weak bound:
\beq
F'(vv'=1) \geq -6.
\eeq
Including the effects of the $\Upsilon$ states below $\bar BB$ thresholds
by assuming that the $\Upsilon \bar BB$ couplings are
of the order of 1, the
bound becomes stronger:
\beq
F'(vv'=1) \geq -1.5.
\eeq
Using QSSR, we can estimate the part
of these couplings entering in the elastic form factor.
We obtain the value of their sum \cite{SN7}:
\beq
\sum g_{\Upsilon \bar BB} \simeq 0.34 \pm 0.02.
\eeq
In order to be conservative, we have multiplied the previous estimate
by a factor 3 larger. We thus obtained the improved bound
\beq
F'(vv'=1) \geq -1.34,
\eeq
but the gain over the previous one is not much. Using the
relation of the form factor with the slope of the Isgur--Wise function,
which differs by $-16/75 \log \alpha_s (M_b)$ \cite{FALK},
one can deduce the final bound:
\beq
\zeta'(1) \geq -1.04.
\eeq
However, one can also use the QSSR expression of the Isgur--Wise function
from vertex sum rules \cite{NEU} in order to extract
the slope $analytically$. To leading order in 1/M,
the $physical$ IW function reads:
\bea
\zeta_{phys}(y\equiv vv')&=& \ga \frac{2}{1+y} \dr ^2
\Bigg \{
1 +\frac{\alpha_s}{\pi}
f(y) \nnb \\
&-&\la \bar dd \ra \tau^3 g(y) +
\la \alpha_sG^2 \ra \tau^4 h(y) \nnb \\
&+&g\la \bar dGd \ra \tau^5 k(y) \Bigg \},
\nnb \\
&&
\eea
where $\tau$ is the Laplace sum rule variable
and $f,~ h$ and $k$ are analytic functions of $y$. From this expression, one
can derive the analytic form of the slope \cite{SN7}:
\beq
\zeta'_{phys}(y=1) \simeq -1 + \delta_{pert} + \delta_{NP},
\eeq
where at the $\tau$-stability region:
$
\delta_{pert} \simeq -\delta_{NP} \simeq -0.04,
$
which shows the near-cancellation of the non-leading corrections.
Adding a generous $50 \%$
error of 0.02 for the correction terms, we finally deduce
to leading order result in 1/M:
\beq
  \zeta'_{phys}(y=1) \simeq -1 \pm 0.02,
\eeq
Using this result in different existing model parametrizations,
we deduce the value of the mixing angle:
\bea
V_{cb} &\simeq& \ga\frac{1.48~\mbox{ps}}{\tau_b}\dr^{1/2}\times
\nnb \\ &.&(37.3 \pm 1.2 \pm 1.4)\times 10^{-3},
\eea
where the first error comes from the data and the second one from the
model dependence.

\nin
 Let us now discuss the effects due to the
$1/M$ corrections. It has been argued recently (but the situation
is still controversial \cite{VAIN}) that the $1/M^2$ effect
can lower the Isgur--Wise function to  $0.91 \pm 0.03$
 at $y=1$, which is a compromise value between the ones in \cite{VAIN},
 such
that the extracted value of $V_{cb}$ using an extrapolation
until this particular point will
also increase by 11$\%$. However, the data from different groups near
this point are very inaccurate and lead to an inaccurate, though
a model-independent result.
 Moreover, in order to see the effect of the $1/M$ correction,
 one can combine this previous result at $y=1$ with the sum rule estimate
of the relevant form factor at $q^2=0$, which is about $0.53 \pm 0.09$
\cite{SN5}, just on top of the CLEO data \cite{ROSS}. Notice that this
result has been obtained without doing a 1/M expansion.
With these two extremal boundary conditions and using the linear
parametrization, which also agrees with the data \cite{ROSS}:
\beq
\zeta = \zeta_0 +\zeta'(y-1),
\eeq
we can deduce the
slope:
\beq
 \zeta' \simeq -( 0.76 \pm 0.2).
\eeq
 It indicates that the $1/M$
correction tends also to decrease $\zeta'$, which implies that, for
larger values of $y$ where the data are more accurate,
the increase of $V_{cb}$ is weaker (+ 3.7$\%$) than the
one at $y=1$. This leads to the $final$ estimate:
\bea
V_{cb} &\simeq& \ga\frac{1.48~\mbox{ps}}{\tau_b}\dr^{1/2}
\times \nnb \\
&.&(38.8 \pm 1.2 \pm 1.5 \pm 1.5)\times 10^{-3}, \nnb \\
&&
\eea
where the new last error is induced by the error from the slope.
This result is more precise than the one obtained at $y=1$, while the
model-dependence only brings a relatively small error.
It also shows that the value from the exclusive channels
is lower than that from the inclusive one,
 which is largely
affected by the large uncertainty
in the mass definition which enters in its fifth power.
Previous results for the slope and for $V_{cb}$ are in good
agreement with the new CLEO data presented at this meeting \cite{ROSS}.
\section{ Properties of hybrid and $B_c$-like mesons}
Let me conclude this talk by shortly discussing the masses of the
hybrid $\bar QGQ$ and the mass and decays of the $B_c$like-mesons.
Hybrid mesons are interesting because of
 their exotic quantum numbers. Moreover,
it is not clear if these states are true resonances or if they only
manifest themselves
as a wide continuum instead. The lowest $\bar cGc$ states appear to be a
$1^{+-}$ of mass around 4.1 GeV \cite{SNB}.
The available sum-rule analysis of
the $1^{-+}$ state is not very conclusive due to the absence of stability
for this channel. However, the analysis indicates
that the spin-one states are in the range 4.1--4.7 GeV.
Their characteristic
decays should occur via the $\eta'~ U(1)$-like particle
produced together with a
$\psi$ or an $\eta_c$. However, the phase-space
suppression can be quite important for these reactions.
The sum rule predicts
that the $0^{--},~0^{++}$ $\bar c Gc$ states are in
the range 5--5.7 GeV, i.e.
about 1 GeV above the spin one.

\nin
We have estimated the $B_c$-meson mass and coupling by combining the results
from potential models and QSSR \cite{BC}. We predict from
potential models:
\bea
M_{B_c}&=& (6255 \pm 20) ~\mbox{MeV},\nnb \\
M_{B^*_c}&=&(6330 \pm 20)~\mbox{MeV},\nnb  \\
M_{\Lambda(bcu)}&=& (6.93\pm 0.05)~\mbox{GeV},\nnb \\
M_{\Omega(bcs)}&=&(7.00 \pm 0.05)~\mbox{GeV}, \nnb \\
M_{\Xi^*(ccu)}&=&(3.63 \pm 0.05)~\mbox{GeV},\nnb \\
M_{\Xi^*(bbu)}&=&(10.21 \pm 0.05)~\mbox{GeV},
\eea
which are consistent with, but more precise than,
the sum-rule results. The decay constant of the $B_c$
meson is better determined from QSSR. The average of the sum rules with the
potential model results reads:
\beq
f_{B_c} \simeq (2.94 \pm 0.12)f_\pi ,
\eeq
which leads to the leptonic decay rate into $\tau \nu_\tau $ of
about $(3.0 \pm 0.4)\times (V_{cb}/0.037)^2 \times 10^{10}~ \mbox{s}^{-1}$.
We have also studied the semileptonic decay of the $B_c$ mesons
and the $q^2$-dependence of the form factors.
We found that, in all cases, the QCD predictions
increase faster than the usual pole dominance ones.
The behaviour can be fitted with an
effective pole mass of about 4.1--4.6 GeV
instead of the 6.3 GeV expected from a pole model.
Basically, we also found that
each exclusive channel has almost
the same rate which is about 1/3 of the leptonic
one. Detection  of these particles in the next
$B$-factory machine will serve
as a stringent
test of the results from the potential models and QSSR
analysis.
\section{Conclusion}
We have shortly presented different results from QCD spectral sum rules
in the heavy-quark sector,
which are useful for further theoretical studies
and complement the results from
lattice calculations or/and heavy-quark symmetry.
{}From the experimental point of view,
QSSR predictions agree with available data,
but they also lead to some new features,
 which need to be tested in forthcoming
experiments.


\end{document}